\documentclass[11pt,twoside]{article}
\usepackage{asp2010}
\usepackage{epstopdf}

\usepackage{hyperref}

\resetcounters

\begin{document}

\title{Long-term stellar activity variations of stars from the HARPS M-dwarf sample: Comparison between activity indices}
\author{J. Gomes da Silva$^{1,2}$, N.C. Santos$^{1,2}$, and X. Bonfils$^{3}$}
\affil{$^1$Centro de Astrof\'isica, Universidade do Porto, Rua das Estrelas, 4150-762 Porto, Portugal}
\affil{$^2$Departamento de F\'isica e Astronomia, Faculdade de Ci\^encias da Universidade do Porto, Portugal}
\affil{$^3$UJF-Grenoble 1 / CNRS-INSU, Institut de Plan\'etologie et d'Astrophysique de Grenoble (IPAG) UMR 5274, Grenoble, F-38041, France}

\begin{abstract}
We used four known chromospheric activity indicators to measure long-term activity variations in a sample of 23 M-dwarf stars from the HARPS planet search program. We compared the indices using weighted Pearson correlation coefficients and found that in general (i) the correlation between $S_{CaII}$ and \ion{Na}{i} is very strong and does not depend on the activity level of the stars, (ii) the correlation between our $S_{CaII}$ and H$\alpha$ seems to depend on the activity level of the stars, and (iii) there is no strong correlation between $S_{CaII}$ and \ion{He}{i} for these type of stars.
\end{abstract}

\section{Stellar magnetic cycles of M-dwarf stars}
There is currently a focus on the search for planets orbiting around M-dwarf stars. Due to their low mass, it will be easier to find lower mass planets orbiting these stars with the radial-velocity (RV) technique. And therefore it is extremely important to access all sources of intrinsic noise that can degrade the quality of the detected RV signals.
There are hints that the magnetic cycles of stars may induce RV capable of hiding the signals of extra-solar planets \citep[e.g.][]{dumusque2010} but not much is known about the long-term activity variations of M-dwarf stars.

The Mt. Wilson survey showed that many solar-like stars have magnetic cycles similar to that of the Sun \citep{wilson1978,baliunas1995} but only one M dwarf was actually included in the sample. This star, HD\,95735, was shown to have long-term activity variations without a defined cycle. More recent studies however uncovered evidence for the existence of periodicity in the long-term activity of a few M stars. Our closest neighbour, Prox Centauri (dMe 5.5), was found to have a magnetic cycle with a $\sim$440 days period \citep{cincunegui2007a}. \citet{diaz2007b} also found a $\sim$760 days periodicity in the activity of the spectroscopic binary Gl375 (dMe 3.5). More recently, \citet{buccino2010} announced the detection of cycles with periods of $\sim$4 and $\sim$7 years for Gl299A (M1/2) and Gl752A (M2.5), respectively.
In a recent paper, we studied the influence of the long-term activity cycles in the RV signals of a sample of 7 early-G and early-K stars known to have activity cycles \citep{santos2010}. We found no hints of RV induced variations by the activity cycles of these stars at the $\sim$1 m s$^{-1}$ level achieved by HARPS. In the present work we extend this study to the lower end of the main sequence by first analyzing the long-term behavior of the chromospheric activity indices and posteriorly compare them with the RV and parameters of the cross-correlation function of the stars.

\section{Sample}
We used a sample of stars from the HARPS M dwarf planet search program \citep{bonfils2010}. This corresponds to a volume limited selection of stars closer than 11 pc, with a declination $\delta < +20$ degrees, brighter than V = 14 mag and with a projected rotational velocity $v\sin i \leq 6.5$ km\,s$^{-1}$. The data consists of high resolution spectra which span around 6 years, from 2003 to 2009.

Our first selection was to remove all spectra with S/N $<$ 2 at the spectral order 6 in the region of the \ion{Ca}{ii} K line. All data was then nightly averaged  and stars with more than 15 measurements were selected. We made bins of 150 days to average out possible rotationally modulated activity variations. For the rest of the study only stars with more than four bins were considered. This process resulted in a sample of 23 stars ranging in spectral type from M0\,V to M4\,V.

\section{Measuring the activity indices}
To measure the activity variations of the stars we computed four known chromospheric activity proxies based on the \ion{Ca}{ii} H and K, H$\alpha$, \ion{Na}{i} D1 and D2, and \ion{He}{i} D3 lines. All these indices were calculated similarly to the procedure used in \citet{santos2000} and \citet{boisse2009}.

Our $S_{CaII}$ index was computed by measuring the flux in the core of the H ($\lambda$3968.47) and K ($\lambda$3933.66) lines using bands of 0.6 \AA~and dividing it by two reference bands of 20 \AA~centered at 3900 and 4000 \AA. This index was then weighted by the square of the errors in each line taken as $\sqrt{N}$ where $N$ is the number of counts inside each band. 
The H$\alpha$ index was computed in a similar fashion, by measuring the flux in a 1.6 \AA~band centered at the line core ($\lambda$6562.808) and dividing it by two reference windows of 10.75 and 8.75 \AA~centered at 6550.87 and 6580.31 \AA, respectively.
We also used a index based on the \ion{Na}{i} $\lambda$5895.92 and $\lambda$5889.95 lines similar to the one proposed by \citet{diaz2007a}. The flux in the cores of the two lines was measured using 0.5 \AA~bands and divided by two reference regions of width 10 and 20 \AA~centered at 5805.0 and 6090.0 \AA, respectively.
Finally, the \ion{He}{i} index was obtained by integrating the flux in a 0.4 \AA~band at the line core at 5875.62 \AA~and dividing it by 5 \AA~reference bands at 5869.0 and 5881.0 \AA~as in \citet{boisse2009}. 

The errors in the four indices were estimated by differentiating the respective equations and taking into account the flux in each band used.

\section{Correlations between the activity indices}
We used the weighted Pearson correlation to compare our four different indices. All coefficients were calculated using the binned data. To access the significance of the obtained coefficients we computed the false-alarm-probability (FAP) by bootstraping the nightly averaged measurements, then binning the data and recalculating the correlation coefficient for each of the 10 000 permutations. We chose significant FAPs those values smaller than 0.05, corresponding to a 95\% significance level. The results are presented in Table \ref{table:corr} with the significant FAPs marked in bold.

\begin{table}[!ht]
\caption{Weighted Pearson correlation between $S_{CaII}$ and the other activity indices. FAPs calculated using bootstrap permutations (see text). On bold are FAP values lower than 0.05.}
\label{table:corr}
\begin{center}
{\small
\begin{tabular}{lccccccc}
\tableline
\noalign{\smallskip}
\multicolumn{1}{l}{Star} &
\multicolumn{1}{c}{$\rho(S_{CaII},\hbox{H}\alpha)$} &
\multicolumn{1}{c}{FAP} &
\multicolumn{1}{c}{$\rho(S_{CaII},$ Na I$)$} &
\multicolumn{1}{c}{FAP} &
\multicolumn{1}{c}{$\rho(S_{CaII},$ He I$)$} &
\multicolumn{1}{c}{FAP} \\
\noalign{\smallskip}
\tableline
\noalign{\smallskip}
GJ361	&	0.127	&	0.4326		&	0.744	&	0.1335		&	-0.086	&	0.4686 \\
Gl1		&	-0.048	&	0.3996		&	0.965	&	\textbf{0.0076}	&	0.611	&	0.1881 \\
Gl176	&	0.932	&	\textbf{0.0036}	&	0.413	&	0.2206		&	0.804	&	\textbf{0.0211} \\
Gl205	&	0.946	&	\textbf{0.0063}	&	0.861	&	\textbf{0.0167}	&	0.267	&	0.3089 \\
Gl273	&	-0.119	&	0.4432		&	0.905	&	\textbf{0.0236}	&	0.386	&	0.2757 \\
Gl382	&	0.997	&	\textbf{0.0001}	&	0.987	&	\textbf{0.0012}	&	0.679	&	0.1162 \\
Gl433	&	-0.407	&	0.2505		&	0.985	&	\textbf{0.0006}	&	-0.370	&	0.3275 \\
Gl436	&	-0.138	&	0.3881		&	0.860	&	\textbf{0.0124}	&	0.436	&	0.1706 \\
Gl479	&	0.992	&	\textbf{0.0040}	&	0.945	&	\textbf{0.0245}	&	0.880	&	0.0633 \\
Gl526	&	-0.823	&	0.1031		&	0.992	&	\textbf{0.0104}	&	0.243	&	0.3520 \\
Gl581	&	-0.656	&	\textbf{0.0176}	&	0.653	&	\textbf{0.0269}	&	0.593	&	0.0624 \\
Gl588	&	0.672	&	0.1179		&	0.953	&	\textbf{0.0173}	&	0.820	&	0.0990 \\
Gl667C	&	-0.512	&	0.0829		&	0.863	&	\textbf{0.0012}	&	-0.487	&	0.0660 \\
Gl674	&	0.411	&	0.3113		&	0.893	&	\textbf{0.0493}	&	0.640	&	0.1998 \\
Gl680	&	0.919	&	\textbf{0.0383}	&	0.643	&	0.1461		&	0.754	&	0.1492 \\
Gl699	&	0.035	&	0.5079		&	0.675	&	0.2239		&	-0.334	&	0.3521 \\
Gl832	&	0.519	&	0.1411		&	0.976	&	\textbf{0.0003}	&	0.251	&	0.3430 \\
Gl849	&	0.695	&	0.0951		&	0.723	&	0.0557		&	0.209	&	0.3733 \\
Gl876	&	0.843	&	0.0984		&	1.000	&	$\mathbf{< 1 . 10^{-5}}$	&	0.771	&	0.1343 \\
Gl887	&	0.831	&	0.1112		&	0.896	&	0.0596		&	0.571	&	0.2291 \\
Gl908	&	-0.512	&	0.1272		&	0.927	&	\textbf{0.0056}	&	0.331	&	0.2973 \\
HIP12961	&	0.383	&	0.2153		&	0.092	&	0.4436		&	0.700	&	0.0738 \\
HIP85647	&	0.955	&	\textbf{0.0112}	&	0.987	&	\textbf{0.0010}	&	0.914	&	\textbf{0.0235} \\
\noalign{\smallskip}
\tableline
\end{tabular}
}
\end{center}
\end{table}

\subsection{$S_{CaII}$ versus H$\alpha$}
\begin{figure}[!ht]
\plotone{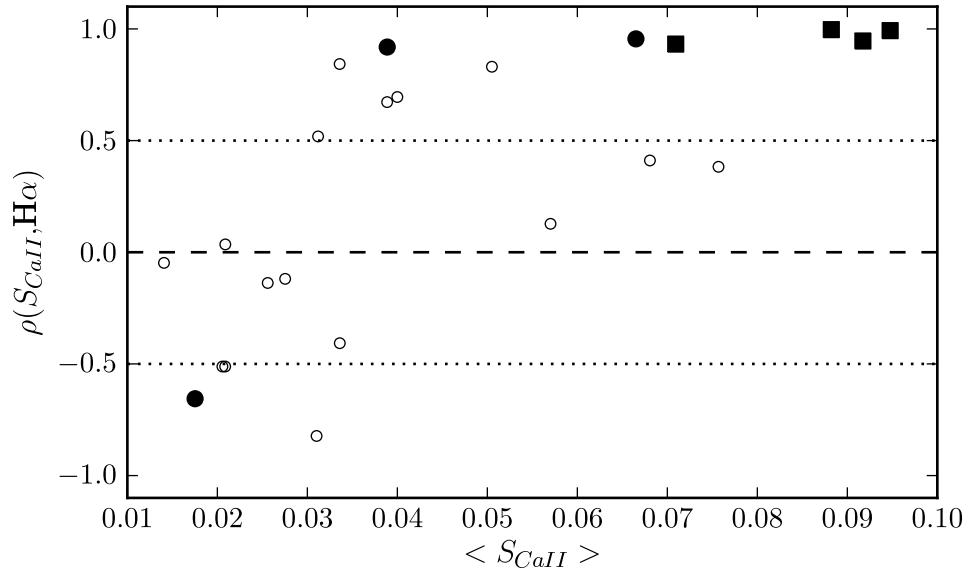}
\caption{Weighted Pearson correlation coefficient between $S_{CaII}$ and H$\alpha$ versus mean activity level. Black squares with FAP $<$ 0.01, black points with 0.01 $<$ FAP $<$ 0.05, small open circles are correlation coefficients with FAP $>$ 0.05.}
\label{fig1}
\end{figure}

From our sample of 23 stars, 7 have significant correlations between the $S_{CaII}$ and the H$\alpha$ index, representing $\sim$30 \% of the total (Table \ref{table:corr}). These stars are Gl176, Gl205, Gl382, Gl479, Gl581, Gl680, and HIP85647. One of them, Gl581, presents anti-correlation between these indices. As was found by \citet{cincunegui2007b} for their study of 109 stars ranging from F6 to M5, we found a great range of correlations with coefficients from $\rho = -0.823$ (Gl526) to $\rho = 0.997$ (Gl382). Figure \ref{fig1} shows what seems to be a trend between the correlation coefficient and the average $S_{CaII}$ for each star. Stars with smaller average activity level tend to have negative correlations while for values of $S_{CaII} > 0.035$ the correlations are all positive.

\subsection{$S_{CaII}$ versus \ion{Na}{i}}
\begin{figure}[!ht]
\plotone{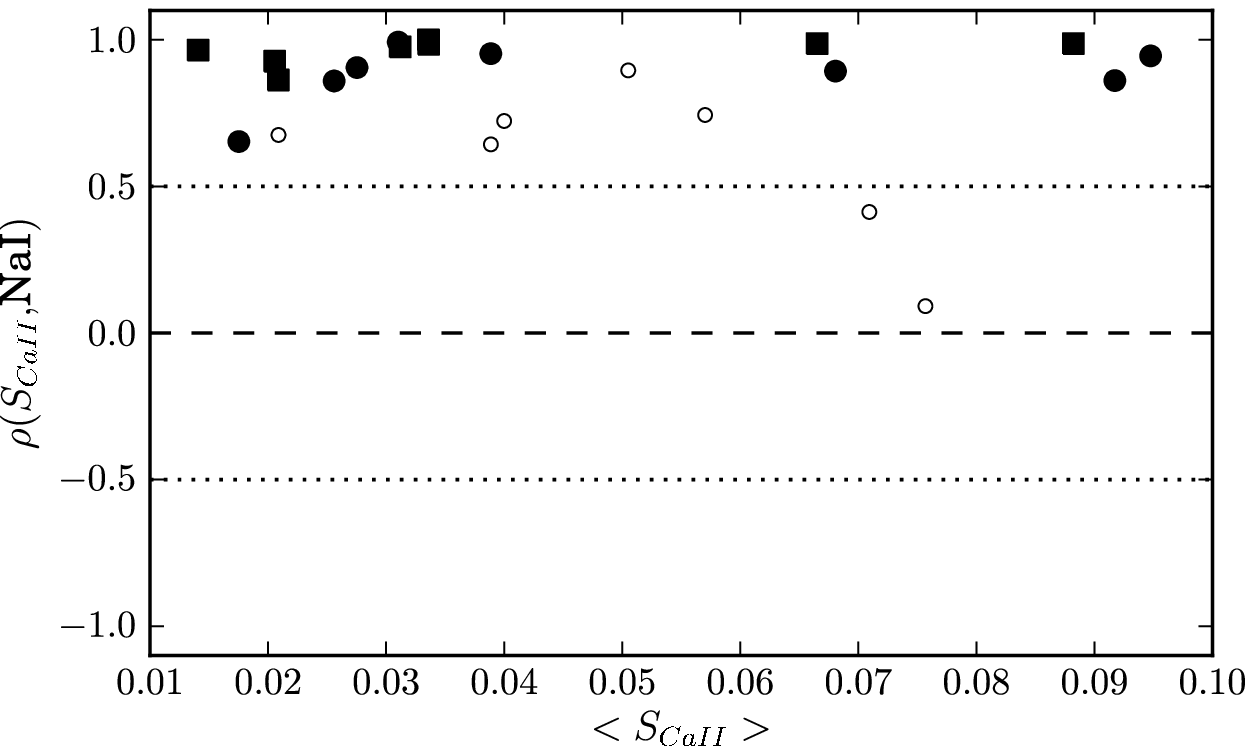}
\caption{Weighted Pearson correlation coefficient between $S_{CaII}$ and \ion{Na}{i} versus mean activity level. Symbols as in Fig. \ref{fig1}.}
\label{fig2}
\end{figure}

Figure \ref{fig2} shows the weighted Pearson correlation coefficient between $S_{CaII}$ and \ion{Na}{i} versus the mean activity level measured by the \ion{Ca}{ii} lines. It is notorious the strong correlations that can be observed between these two indices. There are no negative coefficients and all but two stars (Gl176 and HIP12961) have coefficients with values higher than 0.5. Furthermore, $\sim$70\% of the stars show correlations with FAPs lower than 0.05 and there seems to be no trend between the correlations and the average $S_{CaII}$ activity level. The stars with significative correlation coefficients (FAP $<$ 0.05) are Gl1, Gl205, Gl273, Gl382, Gl433, Gl436, Gl479, Gl526, Gl581, Gl588, Gl667C, Gl674, Gl832, Gl876, Gl908, and HIP85647.

\subsection{$S_{CaII}$ versus \ion{He}{i}}
\begin{figure}[!ht]
\plotone{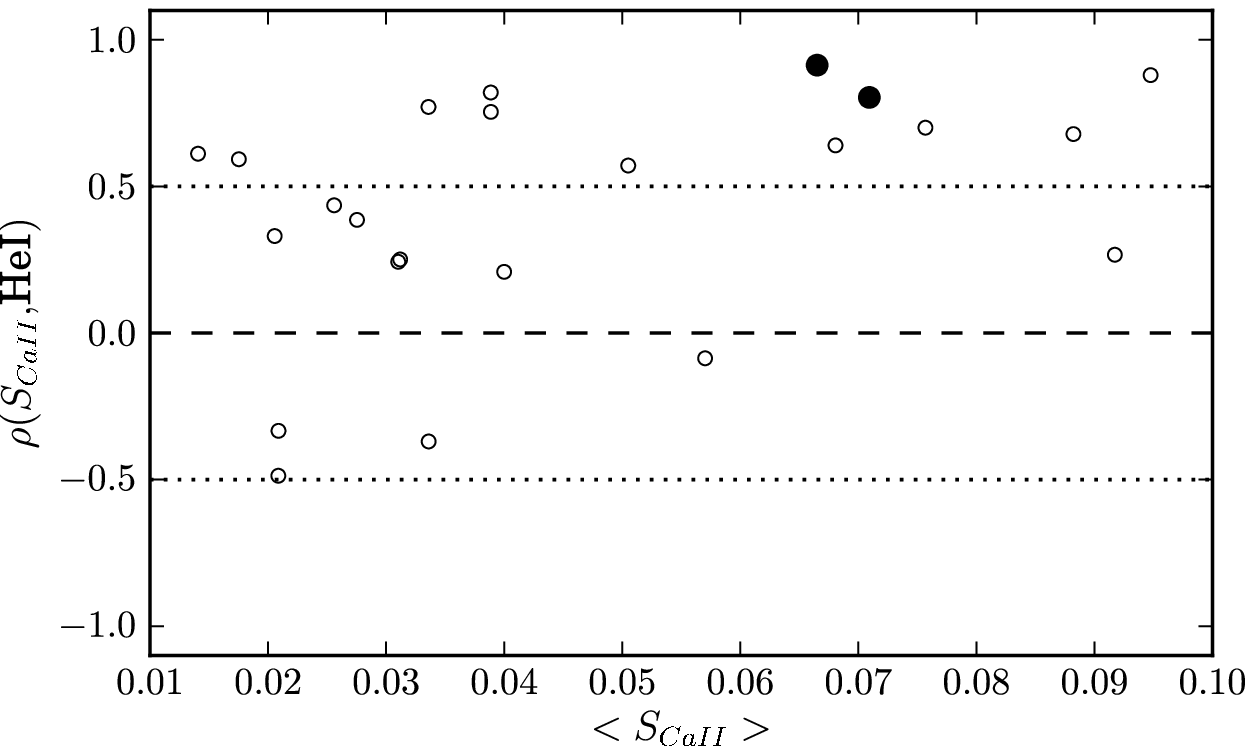}
\caption{Weighted Pearson correlation coefficient between $S_{CaII}$ and \ion{He}{i} versus mean activity level. Symbols as in Fig. \ref{fig1}.}
\label{fig3}
\end{figure}

The correlation between these two indices is very week. As we can observe in Figure \ref{fig3} there is a tendency for positive correlations as only four stars present negative coefficients. But only two stars (Gl176 and HIP85647, representing $\sim$9\% of the sample) have significant correlations between the two indices (with FAP $<$ 0.05). Furthermore, there is a large dispersion of values of the correlation coefficient ranging from $\rho = -0.487$ (Gl667C) to $\rho = 0.914$ (HIP85647).

\section{Conclusions}
We measured four activity indices for a sample of 23 M stars from the HARPS planet search program during a timespan of around 6 years. We compared the activity indices using weighted Pearson correlation coefficients and found that:

\begin{itemize}
\item There is a strong correlation between our $S_{CaII}$ and the \ion{Na}{i} indices. This confirms that the \ion{Na}{i} lines are good activity proxies for these cool stars as suggested by \citet{diaz2007a} for very active stars.
\item As observed by \citet{cincunegui2007b} we found a great range of correlations between the $S_{CaII}$ and H$\alpha$ indices. Furthermore we found what appears to be a trend between the correlation and the average activity level of the stars as measured by the $S_{CaII}$ index.
\item Although some authors suggest the use of the \ion{He}{i} line as a chromospheric activity proxy \citep[e.g.][]{saar1997b} we found that this index is not well correlated with $S_{CaII}$ for M dwarfs.
\end{itemize}

Since the signal-to-noise ratio in the \ion{Ca}{ii} H \& K lines is very low for M dwarfs, we suggest the use of the \ion{Na}{i} D1 and D2 lines, situated in a redder spectral region, as an alternative chromospheric indicator for this type of stars.
These results may influence the way chromospheric activity is accessed in M-dwarf stars and contribute to the knowledge about the activity cycles of such stars.

A more detailed study about this subject will be described in a future publication. Those results will then be used in the context of planet detection to search for trends between the long-term magnetic activity, RV, and parameters of the cross-correlation function in order to access at which level activity cycles might be inducing RV variations for these type of stars.

\acknowledgements This work has been supported by the European Research Council/European Community under the FP7 through a Starting Grant, as well as in the form of a grant reference PTDT/CTE-AST/098528/2008, funded by Funda\c{c}\~ao para a Ci\^encia e a Tecnologia (FCT), Portugal. J.G.S. would like to thank the financial support given by FCT in the form of a scholarship, namely SFRH/BD/64722/2009. N.C.S. would further like to thank the support from FCT through a Ci\^encia 2007 contract funded by FCT/MCTES (Portugal) and POPH/FSE (EC).

\bibliography{gomesdasilva_j}

\end{document}